\documentclass[iopart,superscriptaddress,showpacs,twocolumn]{revtex4}
\usepackage[T1]{fontenc}
\usepackage[latin1]{inputenc}
\usepackage{amsmath}
\usepackage{amssymb}
\usepackage{dsfont}
\usepackage{color}
\makeatletter
\usepackage{hyperref}
\usepackage{amssymb}
\usepackage{graphicx}
\usepackage{epstopdf}
\usepackage{epsfig}
\usepackage{subfigure}
\makeatother

\newcommand{\ket}[1]{|#1\rangle}
\newcommand{\bra}[1]{\langle#1|}

\newcommand{\C}[1]{\hat{#1}^{\dagger}}
\newcommand{\De}[1]{\hat{#1}}

\usepackage{cancel,ifthen}
\usepackage[normalem]{ulem}

\newcommand{\comment}[2][NoInPuT]{\ifthenelse{\equal{#1}{NoInPuT}}{}{{\color{blue}\sout{#1}}}{\color{red} #2}}
\begin{document}

\title{On the Josephson effect in a Bose-Einstein condensate\\ subject to a density dependent gauge potential}

\author{M. J. Edmonds}
\email{mje9@hw.ac.uk}
\affiliation{SUPA, Institute of Photonics and Quantum Sciences, Heriot-Watt University, Edinburgh, EH14 4AS, United Kingdom}
\author{M. Valiente}
\affiliation{SUPA, Institute of Photonics and Quantum Sciences, Heriot-Watt University, Edinburgh, EH14 4AS, United Kingdom}
\author{P. \"Ohberg}
\affiliation{SUPA, Institute of Photonics and Quantum Sciences, Heriot-Watt University, Edinburgh, EH14 4AS, United Kingdom}

\pacs{03.75-b,03.75.Lm,42.50.Gy}
\date{\today{}}

\begin{abstract}\noindent
We investigate the coherent dynamics of a Bose-Einstein condensate in a double well,  subject to a density dependent gauge potential. Further, we derive the nonlinear Josephson equations that allow us to understand the many-body system in terms of a classical Hamiltonian that describes the motion of a nonrigid pendulum with an initial angular offset. Finally we analyze the phase-space trajectories of the system, and describe how the self-trapping is affected by the presence of an interacting gauge potential.    
\end{abstract}
\maketitle
{\it Introduction. }The ability to manipulate individual quantum particles has caused a landslide of fascinating discoveries that have substantially increased our understanding of both the microscopic and macroscopic world. Of principle interest are condensed matter systems where it is now possible to realise Feynman's vision of quantum simulation: the emulation of one system of interest with another\cite{feynman_1982}.   
Atomic condensates formed of bosons and fermions have been utilised to study a plethora of ideas due to the experimental and theoretical control they afford. One prominent example that has been the subject of intense study over the last few years is the ability to create artificial gauge potentials in ensembles of ultracold matter\cite{lin_2009a,lin_2009b,lin_2011}.
The ability to study gauge theories in this context has opened up new directions in the ultracold-atoms landscape such as spin-orbit coupling\cite{edmonds_2012}, Hall physics\cite{leblanc_2012} and even relativistic effects\cite{merkl_2008,ruseckas_2005}. The experimental control achievable over these systems means that one has the ability to study single and many-body physics with a sophisticated level of control.

To create these ``artificial'' vector potentials, one can for instance stir the condensate with a laser, a technique which has been used to realise the vortex lattice of a condensate\cite{aboshaeer_2001}. This has its limitation in that we can only create a constant magnetic field in the rotating frame. Alternately; one can use optical couplings that lead to dark state dynamics\cite{dalibard_2011} or Raman transitions\cite{campbell_2011}. Further, one also has the option to study many-body effects on a lattice. To induce gauge potentials in the discrete setting, laser assisted tunnelling can be used in order to prepare the required phases for the tunnelling amplitudes between the individual sites that constitute the lattice\cite{jaksch_2003,garcia_2012}.

A common feature of both the continuum and the lattice gauge theory is that they are static; in the sense that they are determined by the external laser couplings, and are not affected by the motion of the atoms. It was shown in \cite{edmonds_2012a} how one can induce an effective back action between the gauge field and the dressed states of the light-matter interaction, resulting in a gauge potential that depends on the density of the quantum gas. In this paper we demonstrate how this continuum interacting gauge theory can be applied to a two-site lattice, from which we study the coherent transport between the two sites via the nonlinear Josephson relations which we also derive and numerically solve.

{\it A continuum interacting gauge theory. }The question of how to study and understand the many-body problem lies at the heart of any realistic attempt to construct a theory of interacting particles in many sub-disciplines of physics. It is simply the case that by studying systems of particles with many different, interacting degrees of freedom one is left in a situation that is analytically and numerically intractable. One methodology to tackle this is to partition the particles according the types of motion that occur within the system. The most prominent example is given in atomic systems where the motion of the nucleus and the electrons are divided into slow and fast degrees of freedom respectively, the Born-Oppenheimer approximation.

This dichotomy can be extended to other systems whose degrees of freedom can be separated in a similar fashion. Cold atom systems represent a highly controllable scenario to study many-body physics at ultracold temperatures. However, as these atomic gases are charge neutral, one does not instantly have access to the many paradigmatic effects that pepper the condensed matter physics of charged particles. To redress this, one must find a way to simulate the mathematical structure of a gauge theory with these charge neutral systems.

For bosonic systems, a simple model one can utilize is a system of $N$ particles whose state space is spanned by the atomic states $\{\ket{1},\ket{2}\}$. These states along with the adiabatic theorem give us the ingredients to construct our Born-Oppenheimer like approximation.
Our goal then is to derive an equation of motion for the $N$ two-level systems at the mean-field level. As a first step we define the Hartree wave function for our system as $\ket{\Psi}=\otimes_{i=1}^{N}\ket{\Psi_i}$, where $\ket{\Psi_i}$ is the single particle wave function. Consequentially, one can write down an energy functional that will be the workhorse of the calculation, which is given by $E=\bra{\Psi}\hat{\mathcal{H}}\ket{\Psi}$ with
\begin{equation}
\hat{\mathcal{H}}=\frac{\hat{\bf p}^2}{2m}\otimes\hat{1}+\hat{H}_{lm}+\hat{V}+\hat{\mathcal{V}},
\label{ham0}
\end{equation}
and
\begin{equation}
\hat{H}_{lm}=\frac{\hbar\Omega}{2}\left( \begin{array}{cc}
0& e^{-i\phi({\bf r})}\\
e^{i\phi({\bf r})} & 0\\
\end{array}\right)
\label{lmc}
\end{equation}
is the Hamiltonian for the light-matter interaction. Equation \eqref{ham0} above also contains the important single-particle external potential $\hat{V}$ that we will define later. Bose-Einstein condensates have particle densities that are typically of the order $10^{13}$-$10^{15}$cm$^{-3}$, and as such it is appropriate to model the scattering to leading order by two-body zero range interactions. Hence the interaction matrix $\hat{\mathcal{V}}$ in Eq. \eqref{ham0} is given by $\hat{\mathcal{V}}=(1/2)\mbox{diag}[g_{11}\rho_1+g_{12}\rho_2,g_{22}\rho_2+g_{12}\rho_1]$ and we define the population density in the state $i$ as $\rho_i=|\Psi_i|^2$. To construct an interacting gauge theory, we make use of the dilute property of the gas to construct a perturbation theory using the atomic dressed states of the light-matter coupling Hamiltonian, which we denote $\ket{\chi_{\pm}^{(0)}}=(\ket{1}\pm e^{i\phi}\ket{2})/\sqrt{2}$. We wish to diagonalize $\hat{U}+\hat{\mathcal{V}}$ then by treating the particle interactions as weak compared to the light-matter coupling, so that the chemical potential satisfies $\mu({\bf r})\ll\hbar\Omega$. Consequently, the perturbed dressed states can be written as
\begin{equation}
\ket{\chi_{\pm}}=\ket{\chi_{\pm}^{(0)}}\pm\frac{g_{11}-g_{22}}{8\hbar\Omega}\rho_{\pm}\ket{\chi_{\mp}^{(0)}},
\label{ham_atom}
\end{equation}
and the perturbed spatially varying eigenvalues $g\rho_{\pm}\pm\hbar\Omega/2$ now contain a contribution from the local chemical potential of the gas, the effective scattering parameter being $g=(g_{11}+g_{22}+2g_{12})/4$. We transform the interaction term $\hat{\mathcal{V}}$ in Eq. \eqref{ham0} into the $\pm$ basis by the unitary transformation $\hat{U}^{\dagger}\hat{\mathcal{V}}\hat{U}$, where the transformation between the atomic and dressed basis is given by $\Psi_{l\in\{1,2\}}=\sum_{i=\{+,-\}}\bra{l}\chi_{i}^{(0)}\rangle\Psi_i$. The two-body interaction matrix $\mathcal{V}_{\pm}$ then reads
\begin{equation}
\mathcal{V}_{\pm}=\left(\begin{array}{cc}g & \frac{1}{4}(g_{11}-g_{22}) \\\frac{1}{4}(g_{11}-g_{22}) & g\end{array}\right)\rho_{\pm}.
\end{equation}
To build an interacting gauge theory, we define a state vector comprised of the two basis functions $\ket{\xi}=\sum_{i=+,-}\Psi_{i}\ket{\chi_{i}}$. By projecting onto one of these states we assume the adiabatic theorem is valid, which requires that the un-projected state have negligible population. Thus, the effective Hamiltonian Eq. \eqref{ham0} becomes
\begin{equation}
\hat{H}_{\pm}=\frac{(\hat{{\bf p}}-{\bf A}_{\pm}[{\bf r};\rho_{\pm}({\bf r},t)])^2}{2m}+V_{\pm}({\bf r})+E_{0}+\frac{g}{2}\rho_{\pm},
\label{h_eff}
\end{equation} 
with $E_0=W\pm\hbar\Omega/2$. The density-dependent geometric phase $\textbf{A}_{\pm}=i\hbar\langle\chi_{\pm}|\nabla\chi_{\pm}\rangle$ arises from the spatial dependence of the perturbed dressed states, and is also known as the mead-berry connection\cite{mead_1992}. The scalar geometric phase is defined as $W=\frac{\hbar^2}{2m}|\bra{\chi_{-}}\nabla\chi_{+}\rangle|^2$. Using the definition of $\ket{\chi_{\pm}}$ the vector geometric phase is then given to leading order by ${\bf A}_{\pm}={\bf A}^{(0)}\pm{\bf a}_1\rho_{\pm}({\bf r})$. There is a single as well as a many-body contribution to ${\bf A}_{\pm}$, where the single particle vector potential is ${\bf A}^{(0)}=-\frac{\hbar}{2}\nabla\phi$ and ${\bf a}_1=\nabla\phi(g_{11}-g_{22})/8\Omega$ determines the strength of the density-dependent vector potential.

To study the dynamics of the condensate, we can derive a Gross-Pitaevskii like equation of motion by minimising the energy functional $\mathcal{E}=\bra{\Psi}(i\hbar\partial_t-\hat{H}_{\pm})\ket{\Psi}$. Without loss of generality we minimize with respect to $\Psi_{+}^{*}$, and drop the $\pm$ subscripts on $\rho_{\pm}$, $\Psi_{\pm}$ and ${\bf A}_{\pm}$, thus the mean-field equation of motion reads
\begin{equation}
i\hbar\partial_t\Psi=\bigg[\frac{(\hat{\bf p}-\bf{A})^2}{2m}+{\bf a}_{1}\cdot{\bf j}+V({\textbf{r}})+W+g\rho\bigg]\Psi.
\label{gpe}
\end{equation}
Interestingly, we now have two distinct types of nonlinearity appearing in Eq.\eqref{gpe}, the standard $|\Psi|^2$ term as well as a current ${\bf j}$ that appears at the mean-field level, given by
\begin{equation}
{\bf j}=\frac{\hbar}{2mi}\bigg[\Psi\bigg(\nabla+\frac{i}{\hbar}{\bf A}\bigg)\Psi^*-\Psi^*\bigg(\nabla-\frac{i}{\hbar}{\bf A}\bigg)\Psi\bigg].
\end{equation}
As with the single particle case, the continuity equation that connects the probability density to the probability current is given by $\partial_{t}\rho+\nabla\cdot{\bf j}=0$, although it is stressed that the current in Eq. \eqref{gpe} is a collective effect. The mean-field scalar potential $W$ is given to leading order by $W=|{\bf A}^{(0)}|^2/2m$. This model has recently been studied in a one-dimensional context, where novel effects like chirality\cite{aglietti_1996,edmonds_2012a} are present due to the current {\bf j} and the density-dependent gauge field in Eq. \eqref{gpe}. Experimental realization would require atoms with long lived excited states; for example the transition $^{1}S_{0}\leftrightarrow^{1}P_{1}$ in Sr might be suitable\cite{ye_2008}.

{\it One-dimensional model. }Some of the most striking effects of coherent matter in the ultracold temperature regime have been elucidated with bosonic atomic condensates. Early experiments and theoretical work focussed on understanding interference with matter waves\cite{andrews_1997,javanainen_1996}, Bragg scattering\cite{strenger_1999} and applications to matter wave optics such as the realization of nonlinear effects like solitons\cite{denschlag_2000}. One of the most surprising yet paradigmatic effects in quantum mechanics is the quantum tunnelling of particles. 
For macroscopic systems this is encompassed by the Josephson effect: the tunnelling current that is produced by placing an insulating barrier between two particle reservoirs. The inherent nonlinearities that are present in the study of interacting ultracold bosonic gases make these systems particularly interesting, and has lead to the prediction and realization of effects such as self-trapping\cite{smerzi_1997,albiez_2005,levy_2007}. A detailed overview of these effects is given by Gati et. al. \cite{gati_2007}.
Extensions of the Josephson effect to systems incorporating non-abelian gauge fields have recently been described\cite{merkl_2010,qi_2009}. Here, we show how the interacting gauge theory described in the previous section can be placed onto a two-site lattice, and study the population dynamics of the resulting lattice gauge theory (LGT). 

We envisage the situation where the cloud of atoms is confined so that any transversal dynamics are effectively frozen out, meaning a one-dimensional mean-field description of the BEC is justified. We also define $\phi=kx$ as the phase of the incident laser, and also re-define the state $\Psi(x)=e^{-ikx/2}\psi(x)$, which yields:
\begin{equation}
i\hbar\partial_t\psi=\bigg[\frac{1}{2m}(\hat{p}-a_1\rho)^2+a_1j(x)+V(x)+\tilde{W}+g\rho\bigg]\psi,
\label{gpe1d}
\end{equation}
with $\tilde{W}=\hbar^2k^2/8m$, and $a_1=k(g_{11}-g_{22})/8\Omega S_t$ gives the strength of the current nonlinearity. The parameter $S_t$ defines the transversal area of the 1D cloud. Next, we perform a gauge transformation on Eq. \eqref{gpe1d}, 
\begin{equation}
\psi(x,t)=\exp\bigg\{\frac{ia_1}{\hbar}\int\limits_{-\infty}^{x}dx'\rho(x',t)-i\tilde{W}t/\hbar\bigg\}\Phi(x,t),
\end{equation}
the resulting equation of motion for $\Phi(x,t)$ is given by
\begin{equation}
i\hbar\partial_t\Phi=\bigg[-\frac{\hbar^2}{2m}\partial_{x}^{2}+V(x)-2a_1j(x)+g|\Phi|^2\bigg]\Phi,
\label{gpe1da}
\end{equation}
where the gauge transformed current appearing now in Eq. \eqref{gpe1da} is given by $j(x)=(\hbar/m)\mbox{Im}(\Phi^*(x)\partial_x\Phi(x))$. We can then write the one-dimensional mean-field Hamiltonian in Eq. \eqref{gpe1da} in second-quantized form
\begin{align}\nonumber
\hat{H}=&\int dx \ \C{\Phi}(x)\bigg(-\frac{\hbar^2}{2m}\partial_{x}^{2}+V(x)\bigg)\De{\Phi}(x)\\\nonumber&+g\int dx\ \C{\Phi}(x)\C{\Phi}(x)\De{\Phi}(x)\De{\Phi}(x)\\&-2a_1\int dx\ \C{\Phi}(x)\hat{J}(x)\De{\Phi}(x),
\label{ham}
\end{align}
where the normal ordered current operator $\hat{J}(x)$ that appears in Eq. \eqref{ham} is given in second quantized form by
\begin{equation}
\hat{J}(x)=\frac{\hbar}{2mi}\bigg[\C{\Phi}(x)\partial_x\De{\Phi}(x)-\partial_x\C{\Phi}(x)\De{\Phi}(x)\bigg].
\label{curr}
\end{equation}
To proceed, we require a model potential $V(x)$. Typically one is interested in situations where the tight binding approximation can be made, such that the height of the lattice is greater than the chemical potential at each individual well. Experimentally, an extended one-dimensional lattice can be realized by counter-propagating laser beams and the fact that the energy of an individual atom is shifted by an amount $\Delta E=-\frac{1}{2}\alpha'(\omega)\langle\mathcal{E}({\bf r},t)^2\rangle_t$, where $\alpha'(\omega)$ is the real part of the dynamical polarizability of the atom, and $\langle\mathcal{E}({\bf r},t)^2\rangle_t$ is the time averaged square of the electric field. However, as we are interested in a two-site system, schemes involving electrostatic interactions\cite{kruger_2003} or atom chips\cite{schumm_2005} are more appropriate. Here, we consider the model potential \cite{milburn_1997}
\begin{equation}
V(x)=b\bigg(x^2-\frac{\frac{1}{2}m\omega^2}{2b}\bigg)^2.
\label{pot_dw}
\end{equation}
The double-well potential defined by Eq. \eqref{pot_dw} above has its minima situated at $x_{\text{min}}=\pm\sqrt{m\omega^2/4b}$, and close to these points $V(x)$ is harmonic, an approximation we will use to perform the tight binding calculation. Hence the normalized local ground states of the left and the right well are given by $\eta_{l/r}(x)=(2/\pi\sigma^2)^{1/4}\exp(-[(x\pm x_{\text{min}})/\sigma]^2)$ respectively, and we define $\sigma=\sqrt{2\hbar/m\omega}$ as the width of the ground states.

{\it Mean-field tight binding calculation. }We are now in a position to derive a mean-field tight binding Hamiltonian describing the population dynamics between the two wells. To do this we use the two-mode approximation, which entails expanding the field operator $\De{\Phi}$ as 
\begin{equation}
\De{\Phi}=\eta_l(x)\De{c}_l+\eta_r(x)\De{c}_r,
\label{2mode}
\end{equation} 
where the operator $\De{c}_l$ and $\De{c}_r$ destroy particles in the left and right wells respectively. Note that we are assuming that there is a large separation between the ground state and the first excited state of each individual well so that the dynamics are well described by assuming the particles are located in one of the two ground states. If we insert Eq. \eqref{2mode} into the Hamiltonian given previously by \eqref{ham}, we arrive at:
\begin{align}\nonumber
\hat{H}=&\sum_{ij}J_{ij}\C{c}_i\De{c}_j+\sum_{ijkl}U_{ijkl}\C{c}_i\C{c}_j\De{c}_k\De{c}_l\\&+\sum_{ijkl}\lambda_{ijkl}\C{c}_i\C{c}_j\De{c}_k\De{c}_l,
\label{ham_dis}
\end{align}
the sums being taken over both the left and right wells. The parameters $J_{ij}$, $U_{ijkl}$ and $\lambda_{ijkl}$ are overlap integrals, and give us a way to identify the most important terms in Eq. \eqref{ham_dis}. The overlap integrals can be readily calculated using our definition of the ground states $\eta_{l/r}(x)$ given previously, thus one finds that the lattice Hamiltonian Eq. \ref{ham_dis} simplifies to
\begin{figure} 
\centering
\includegraphics[scale=0.22]{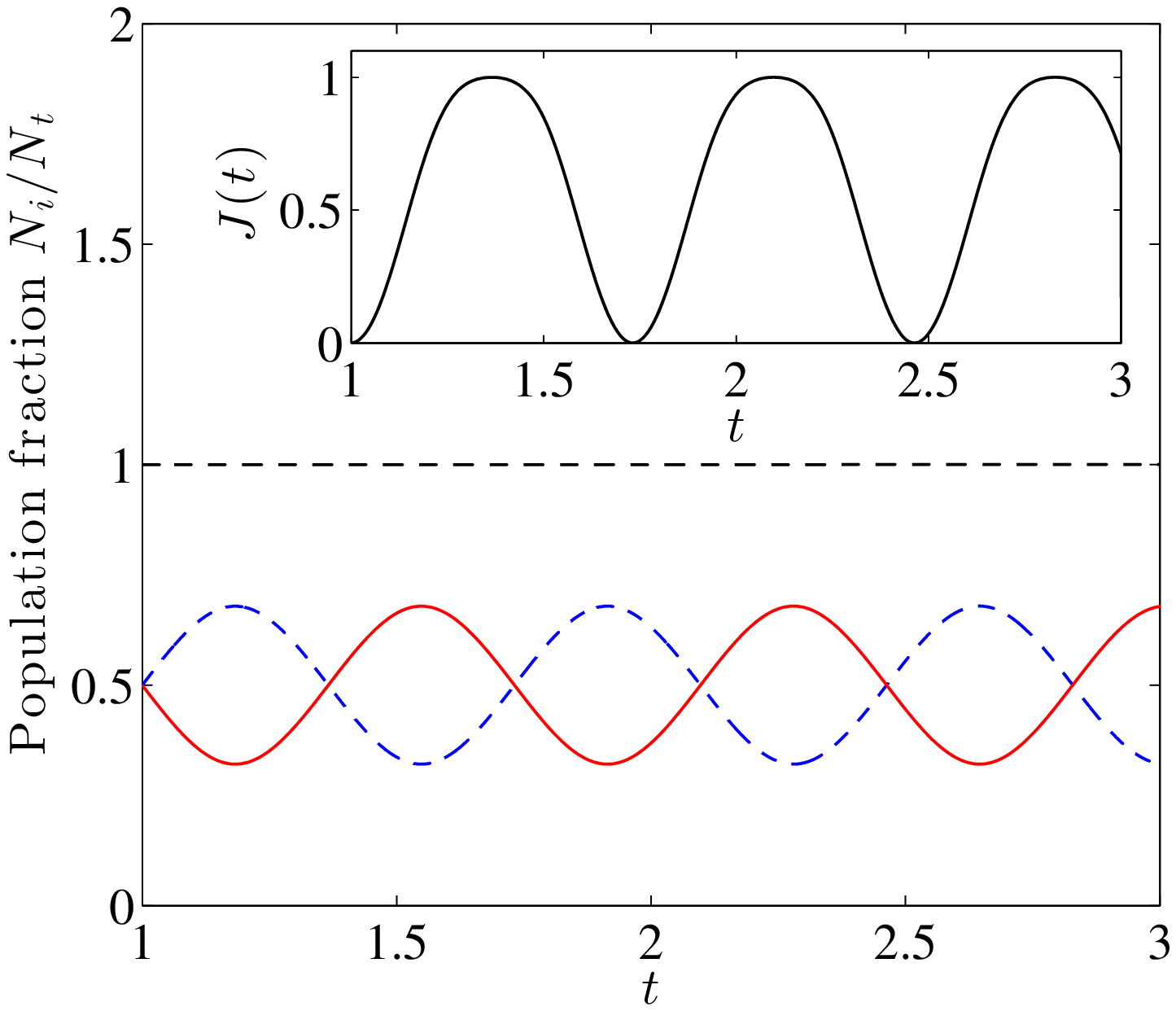}
\includegraphics[scale=0.22]{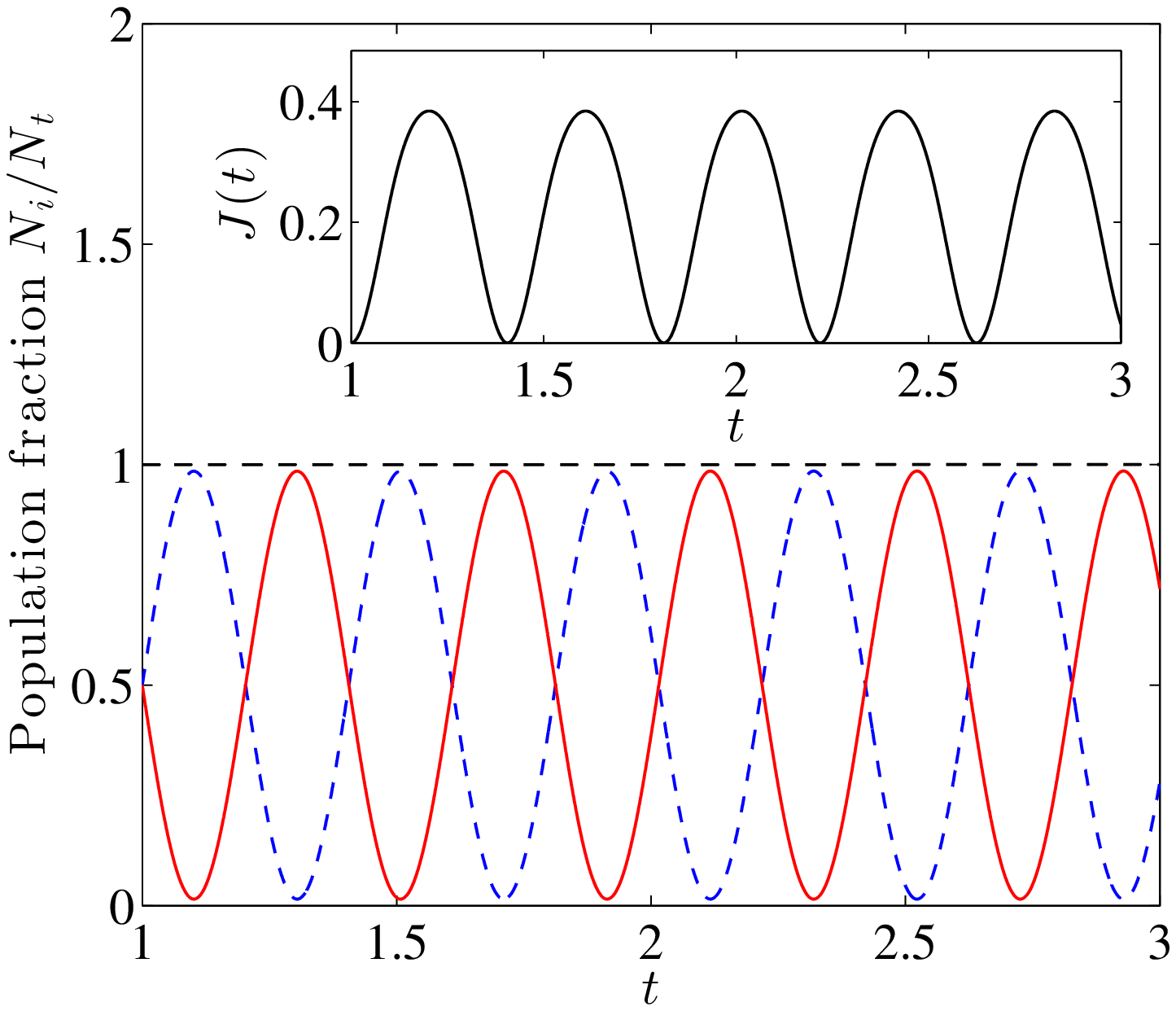}\\
{(a)\hspace{4cm}(b)}\\
\includegraphics[scale=0.22]{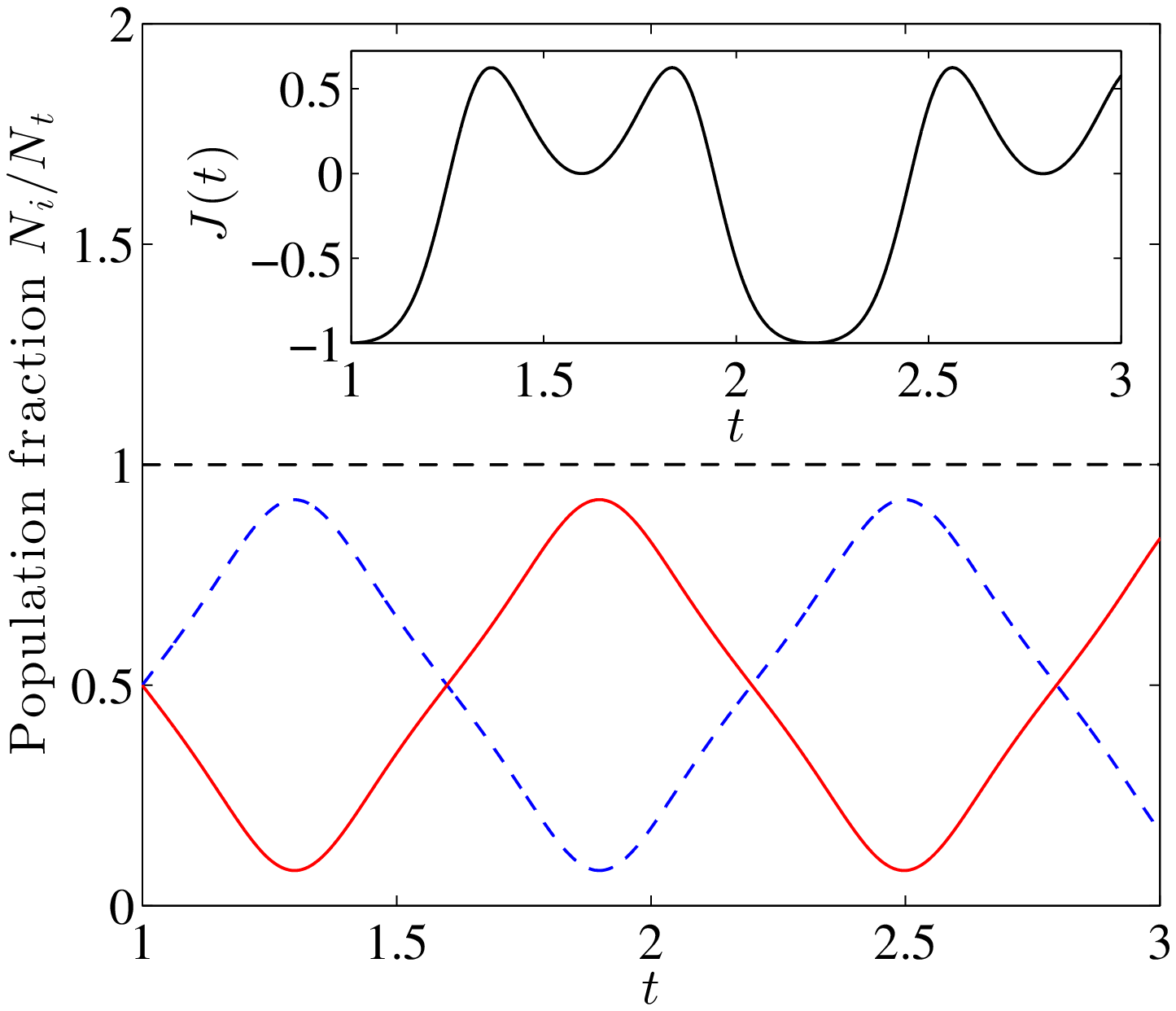}
\includegraphics[scale=0.22]{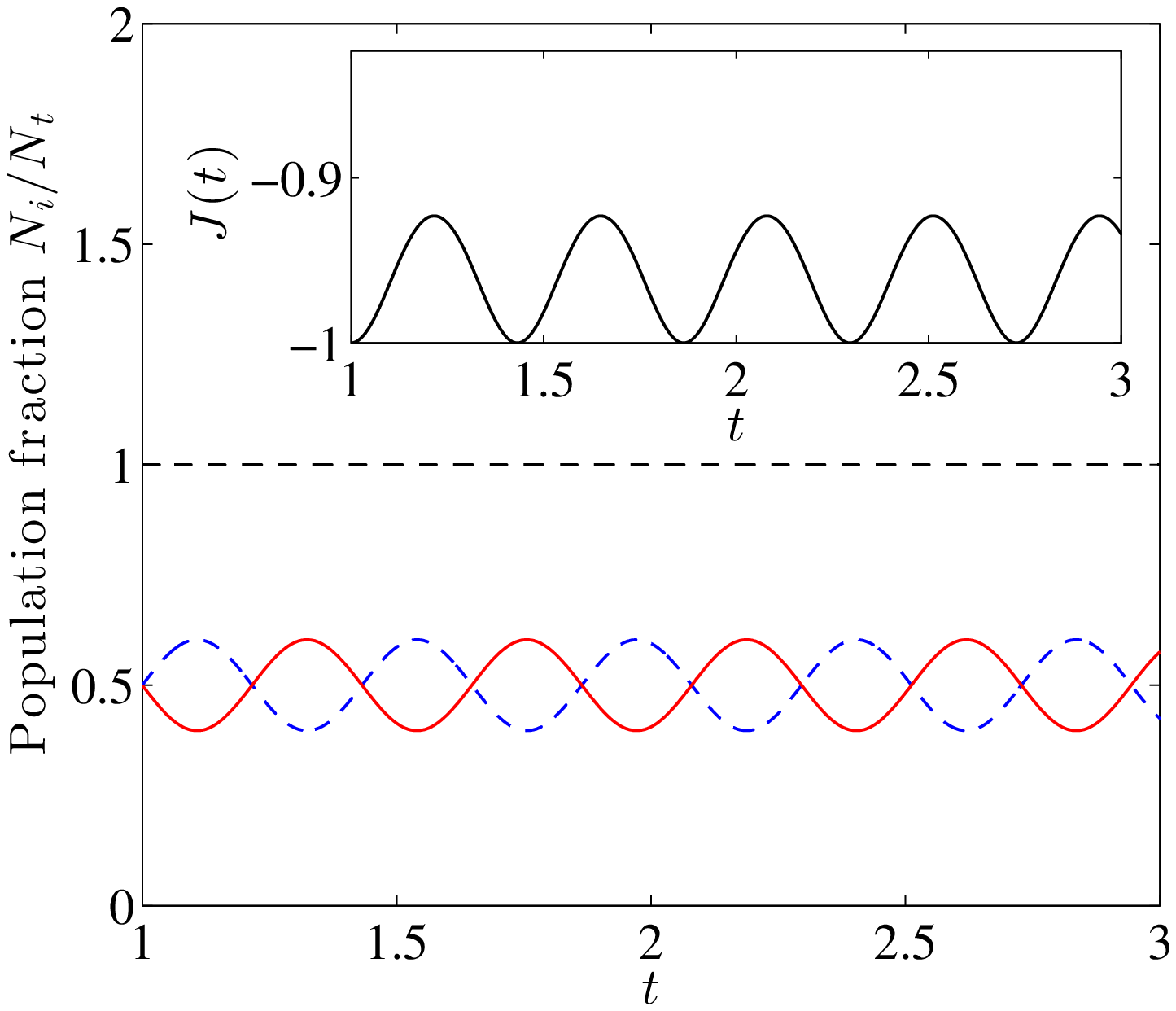}\\
{(c)\hspace{4cm}(d)}
\caption{(Color online) Numerical solutions to Eq. \eqref{heis_cl} and \eqref{heis_cr}. The population $|c_l|^2$, $|c_r|^2$ and $\sum_i|c_i|^2$ are given by the blue dashed line, solid red line and dashed black lines respectively. Figure (a) and (c) show the population oscillations for $2U/J=10$ and $\Gamma_1/J=1$, the insets in each figure show the current $J(t)$ as a function of time. Both (a) and (b) have the initial conditions $c_{l/r}(t=0)=1/\sqrt{2}$. Figure (b) and (d) are plotted for $2U/J=1$ and $\Gamma_1/J=5$. Figures (c) and (d) have the initial conditions $c_{l}(t=0)=e^{i\pi/2}/\sqrt{2}$ and $c_{r}(t=0)=1/\sqrt{2}$. The units of time are $\hbar/J$.}
\label{fig1}
\end{figure}
\begin{align}\nonumber
\hat{H}=&-J(\C{c}_l\De{c}_r+\C{c}_r\De{c}_l)+U(\hat{n}_l(\hat{n}_l-1)+\hat{n}_r(\hat{n}_r-1))\\&+\Gamma_{1}(\C{c}_l\hat{j}\De{c}_l+\C{c}_r\hat{j}\De{c}_r)+\Gamma_{2}(\C{c}_l\hat{j}\De{c}_r+\C{c}_r\hat{j}\De{c}_l),
\label{ham_latt}
\end{align}
where $\hat{n}_i=\C{c}_{i}\De{c}_i$ is the number operator for site $i$ and the constants that appear in Eq. \eqref{ham_latt} are given by $U=g/\sqrt{\pi}\sigma$, $\Gamma_{1}=8\hbar a_1x_{\text{min}}e^{-3x_{\text{min}}^{2}/\sigma^2}/\sqrt{\pi}\sigma^3$, $\Gamma_{2}=8\hbar a_1x_{\text{min}}e^{-4x_{\text{min}}^2/\sigma^2}/\sqrt{\pi}\sigma^3$ and the discrete current operator for the two site lattice we define by $\hat{j}=-i(\C{c}_r\De{c}_l-\C{c}_l\De{c}_r)$. Equation \eqref{ham_latt} gives us a way to understand the effect of population dynamics between the two sites. It comprises the usual on-site interactions that appear in the Bose-Hubbard model given by the terms proportional to $U$, as well as the unconventional terms proportional to $\Gamma_i$ which originate from the current operator, Eq. \eqref{curr}. To study the population oscillations between the two wells we work with the operators $\De{c}_i$ in the Heisenberg picture. The equations of motion are then given by $i\hbar\dot{\De{c}}_i=[\De{c}_i,\hat{H}]$, which gives
\begin{align}\label{heis_cl}\nonumber
i\hbar\frac{d\De{c}_l}{dt}=&-J\De{c}_r+2U\hat{n}_l\De{c}_l+\Gamma_{1}(\hat{j}\De{c}_l+i(\hat{n}_l+\hat{n}_r)\De{c}_r)\\&+\Gamma_2(\hat{j}\De{c}_r+i\C{c}_l\De{c}_r\De{c}_r+i\hat{n}_r\De{c}_l),\\ \nonumber
i\hbar\frac{d\De{c}_r}{dt}=&-J\De{c}_l+2U\hat{n}_r\De{c}_r+\Gamma_{1}(\hat{j}\De{c}_r-i(\hat{n}_l+\hat{n}_r)\De{c}_l)\label{heis_cr}\\&+\Gamma_2(\hat{j}\De{c}_l-i\C{c}_r\De{c}_l\De{c}_l-i\hat{n}_l\De{c}_r).
\end{align}
To gain an understanding of Eq. \eqref{heis_cl} and \eqref{heis_cr} we assume that the number of particles in both wells is so large that the operators $\De{c}_l$ and $\De{c}_r$ may be treated as classical quantities, and as such can be replaced by their expectation values $c_l$ and $c_r$. Figure \ref{fig1} shows the numerical solutions to these equations in different parameter regimes. In Figure 1 (a) and (c) the parameters $2U/J=10$ and $\Gamma_1/J=1$ were used, whilst for (b) and (d) $2U/J=1$ and $\Gamma_1/J=5$. For figures (a) and (b) the initial phase different was $\theta_l-\theta_r=\Delta\theta=0$, but for figure (c) and (d) $\Delta\theta=\pi/2$ was used. We further assumed that $x_{\text{min}}/\sigma\gg 1$, so that $\Gamma_2\ll\Gamma_1$. Figure \ref{fig1} shows the Rabi like oscillations that are synonymous with two state systems. We see in (b) the increased current strength has caused the speed of the population oscillations to increase. On the other hand, Fig. \ref{fig1} (c) shows how strong on-site interactions changes the dynamics, the current showing an unusual `dip' (see inset). Finally Fig. \ref{fig1} (d) shows how the dynamics are reduced when there is an initial phase difference of $\pi/2$ between the sites and the current strength is stronger than the Hubbard term.

{\it Phase-space analysis. }Figure \ref{fig1} shows how the current nonlinearity in Eq. \eqref{ham_latt} affects the population oscillations between the two wells. To investigate the properties of this system further, we can re-cast the variables of the problem in terms of the population difference between the two sites and the phase difference. This methodology has previously been utilised to show how a charge neutral interacting BEC in a symmetric double well potential can be understood in terms of a nonrigid pendulum\cite{smerzi_1997,raghavan_1999}, and gives an intuitive way to study the phase-space properties of the many-body system\cite{mahmud_2005}. We begin by assuming again that the number of particles is so large that the operators that appear in Eq. \eqref{ham_latt} can be substituted for their respective eigenvalues, which in turn can be replaced by the polar variables $c_i=\sqrt{N_i}e^{i\theta_i}$, where $i=l,r$. If we define the population and phase differences by $z(t)=(N_l-N_r)/N_t$ and $\varphi(t)=\theta_r-\theta_l$ respectivly, then we obtain a classical Hamiltonian from Eq. \eqref{ham_latt}        
\begin{align}\nonumber
H=&\frac{\Lambda z^2}{2}-\sqrt{1-z^2}\cos(\varphi)-\gamma_1\sqrt{1-z^2}\sin(\varphi)\\&-\gamma_2(1-z^2)\sin(2\varphi)+\Delta E,
\label{ham_class}
\end{align}
where the dimensionless parameters read $\Lambda=UN_{t}^{2}/J$, $\gamma_i=\Gamma_i N_{t}^{2}/J$ and $\Delta E=UN_{t}^{2}/2J$. The tunnelling current is $j=-N_t\sqrt{1-z^2}\sin(\varphi)$. The variables $z$ and $\varphi$ are canonically conjugate, with $\dot{z}=-\frac{\partial H}{\partial\varphi}$ and $\dot{\varphi}=\frac{\partial H}{\partial z}$, which gives the coupled equations of motion
\begin{align}\nonumber
\dot{z}=&-\sqrt{1-z^2}\sin(\varphi)+\gamma_1\sqrt{1-z^2}\cos(\varphi)\\&+2\gamma_2(1-z^2)\cos(2\varphi), \label{z} \\ \nonumber
\dot{\varphi}=&\frac{z}{\sqrt{1-z^2}}\cos(\varphi)+\Lambda z+\gamma_1\frac{z}{\sqrt{1-z^2}}\sin(\varphi)\\&+2\gamma_2 z\sin(2\varphi). \label{varphi}
\end{align}
The Hamiltonian Eq. \eqref{ham_class} and the nonlinear Josephson equations Eq. \eqref{z} and \eqref{varphi} allow us to understand the dynamics and phase-space properties of the two site model. In particular, we note that Eq. \ref{ham_class} can be understood in terms of a classical nonrigid pendulum. Our model differs from that presented in \cite{smerzi_1997} by the additional term proportional to $\gamma_2$. Hence, if we drop the $\gamma_2$ term from Eq. \eqref{ham_class} we can map the classical Hamiltonian onto the nonrigid pendulum model as presented in \cite{smerzi_1997}. The result is that the phase angle $\varphi$ has an extra initial offset term due to the current nonlinearity of the underlying continuum model. In this limit Eq. \eqref{ham_class} can be simplified to
\begin{figure} 
   \centering
   \subfigure[]{
   \includegraphics[scale=0.45]{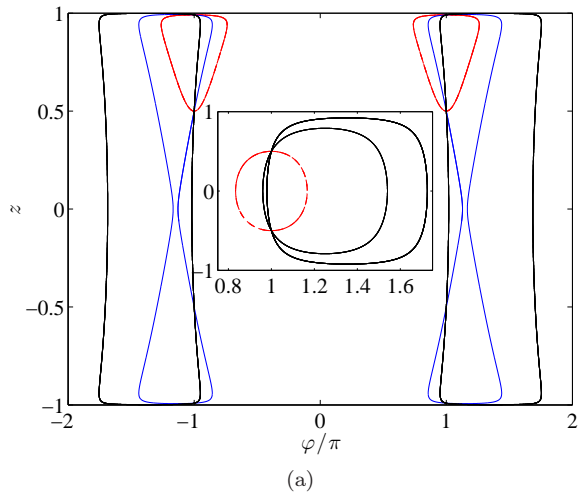}}
   \caption{(Color online) Lines of constant energy, with $z$ vs. $\varphi$. The initial conditions were $z(0)=0.5$ and $\varphi(0)=\pi$. The figure and its inset demonstrate two regimes: the main figure has $\Lambda=2$ and $\gamma_1=\{0,0.5,2\}$. The inset has $\Lambda=0$ and $\gamma_1=\{0,1,2\}$, where the different values of $\gamma_1$ correspond to the smallest through to the largest curves in the figures above.}
   \label{fig2}
\end{figure}
\begin{equation}
H=\frac{\Lambda z^2}{2}-R\sqrt{1-z^2}\cos(\varphi-\varphi_0),
\end{equation}  
where $R=\sqrt{1+\gamma_{1}^{2}}$ and the angle $\varphi_0=\text{arctan}(\gamma_1)$, and we set $\Delta E=0$ without loss of generality. Figure \ref{fig2} shows the phase-space trajectories of variables $z,\varphi$. In Fig. \ref{fig2} we are interested in two limits, the first being the phase-space with on-site interactions, $\Lambda=2$, whilst the inset is plotted for no on-site interactions, $\Lambda=0$. This is justifiable as we could for example use Feshbach resonances in order to achieve $g=0$. The plots for $\Lambda=2$ show how increasing the strength of the current causes the curves obtained from the numerical solutions to equations \eqref{z} and \eqref{varphi} to increase in size. The inset shows how increasing the strength of the current without on-site interactions gives a displacement of the curves by an amount $\varphi_0$. To further quantify this, we calculate the critical value $\Lambda_c$ that determines the point at which the nonrigid pendulum is given an initial kick that pushes it over the vertical $\varphi=\pi$ point. This is given by the condition $H[z(0),\varphi(0)]>1$ so that
\begin{equation}
\Lambda_c=\frac{2(1+\sqrt{1+z(0)^2}R\cos[\varphi(0)-\varphi_0])}{z(0)^2}.
\end{equation}
If we choose $\Lambda=\Lambda_c$ we are in the situation whereby the pendulum is in an upright position, so that the period of oscillation diverges.

{\it Conclusions. }It was shown how a one-dimensional mean-field theory describing density dependent gauge fields can be cast into a two-site tight binding model, using a symmetric double-well potential. It was seen that this discrete formulation also features a current operator as well as the usual on-site interactions that appear in the standard Bose-Hubbard model. Finally, a phase-space analysis was presented, by way of the nonlinear Josephson equations. It was found that a classical Hamiltonian can be written down that describes the motion of a nonrigid pendulum, with an initial angular offset that depends on the strength of the density dependent gauge potential.
Further analysis of this model is warranted. For instance, the unconventional transport mechanisms could have a significant impact on the superfluid behaviour, and is likely to affect non-trivially the elementary excitations of the system.
 
{\it Acknowledgments. }M.J.E. acknowledges support from the EPSRC CM-DTC, and M.V. and P.\"O. from EPSRC grant No. EP/J001392/1.

\end{document}